\documentclass[aps,pre,showpacs,showkeys,twocolumn,preprintnumbers,floatfix,nofootinbib,10pt]{revtex4-2}
\usepackage{eurosym}
\usepackage{amsmath}
\usepackage{amssymb}
\usepackage{graphicx,color}
\usepackage{multirow}

\usepackage[normalem]{ulem}
\usepackage{soul}
\sethlcolor{yellow} 

\usepackage{color}
\usepackage{xcolor}

\begin{document}

\title{ The number of spanning trees as an indicator of critical phenomena:
When Kirchhoff meets Ising}

\author{ Roberto da Silva$^{1}$, Henrique A. Fernandes$^{2}$, Paulo F. Gomes$^{3}$, Sebastian Gon\c{c}alves$^{1}$, E. V. Stock$^{1}$, A. Alves$^{4}$ }
\affiliation{
$^1$Instituto de F{\'i}sica, Universidade Federal do Rio Grande do Sul, Av. Bento Gon{\c{c}}alves, 9500 - CEP 91501-970, Porto Alegre, Rio Grande do Sul, Brazil.\\
$^2$Instituto de Ci{\^e}ncias Exatas e Tecnol{\'o}gicas, Universidade Federal de Jata{\'i}, BR 364, km 192, 3800 - CEP 75801-615, Jata{\'i}, Goi{\'a}s, Brazil.\\
$^3$Faculdade de Ciências e Tecnologia, Universidade Federal de Goiás, Estrada Municipal, CEP 74971-451, Aparecida de Goiânia, Goiás, Brazil.\\ 
$^4$Departamento de F\'isica, Universidade Federal de S\~ao Paulo, UNIFESP, 09972-270, Diadema-SP, Brazil}

\begin{abstract}
Visibility graphs are spatial interpretations of time series. When derived
from the time evolution of physical systems, the graphs associated with such
series may exhibit properties that can reflect aspects such as ergodicity,
criticality, or other dynamical behaviors. It is important to describe how
the criticality of a system is manifested in the structure of the
corresponding graphs or, in a particular way, in the spectra of certain
matrices constructed from them. In this paper, we show how the critical behavior of an Ising spin system manifests in the spectra of the adjacency and Laplacian matrices constructed from an ensemble of time evolutions simulated via Monte Carlo (MC) Markov Chains, even for small systems and short MC steps. In particular,
we show that the number of spanning trees ---or its logarithm---, which represents a kind of \emph{structural entropy} or \emph{topological complexity} here obtained
from Kirchhoff's theorem, can, in an alternative way, describe the criticality of the spin system. These findings parallel those obtained from the spectra of correlation matrices, which similarly encode signatures of critical and chaotic behavior.
\end{abstract}

\maketitle



\textbf{INTRODUCTION:} Capturing physical properties from graphs or networks has been extensively explored in
Statistical Mechanics \cite{Albert2002,Newman2}. Among the many concepts related to these two terms (here, the only difference between them lies in the
level of abstraction and the context of application), the concept of a tree
is particularly important (for an excellent textbook on graph theory, see 
Ref. \cite{Harary1969}). A tree is the simplest way of connecting all nodes in a graph without redundancies, that is, without cycles. In other words, a tree is a minimal structure in the sense that removing any edge would break the connectivity of the graph.

Given a graph, a spanning tree is a connected subgraph that includes all the nodes of the original graph and contains no cycles. The number of spanning trees of a connected graph $G$, denoted by $\tau(G)$, is a fundamental combinatorial quantity that appears in a wide
range of theoretical and applied contexts. This number connects concepts
from graph theory, probability, physics, and information theory.

That number can be regarded as a measure of the \emph{robustness of
connectivity} in a network. Graphs with larger $\tau(G)$ possess more
independent ways of connecting all vertices without forming cycles,
indicating a higher tolerance to structural failures or edge removals.

This idea has been extensively used in network reliability analysis. Boesch
and Myers~\cite{Boesch1984} showed that $\tau(G)$ is directly related to
reliability polynomials, that is, functions measuring the probability that a
graph remains connected when its edges or vertices fail randomly with a
fixed probability, while Colbourn \cite{Colbourn1987} provided a detailed
combinatorial framework for analyzing redundant connectivity in
communication networks.

A classical physical interpretation of $\tau(G)$ arises from its analogy
with \emph{electrical resistor networks}. Kirchhoff's pioneering work~\cite{Kirchhoff1847} established that $\tau(G)$ equals any cofactor of the
Laplacian matrix $L(G)$, an observation known as the \emph{Matrix-Tree
Theorem}. This theorem later became central to the study of network
topology. In an electrical interpretation, Doyle and Snell \cite%
{DoyleSnell1984} demonstrated that random walks and effective resistances in
a network are governed by the Laplacian structure. Klein and Randic \cite{KleinRandic1993} 
further introduced the notion of \emph{resistance distance}, connecting 
$\tau(G)$ to measures of effective conductance in molecular and electrical networks.

Aldous~\cite{Aldous1990} employed random walks to construct uniform spanning
trees, demonstrating that the probability of obtaining a particular tree $T$
is $1/\tau(G)$. Subsequently, Wilson \cite{Wilson1996} proposed a more
efficient random-walk-based algorithm for generating uniform spanning trees,
now known as \emph{Wilson's algorithm}. These developments established a
direct connection between $\tau(G)$, stationary distributions of Markov
chains, and the enumeration of arborescences in directed graphs. 

The logarithm of $\tau(G)$, denoted $\log \tau(G)$, can be viewed as a measure
of \emph{structural entropy} or \emph{topological complexity}. Anand and
Bianconi \cite{Anand2009} formalized this idea within an
information-theoretic framework, showing that networks with larger $\tau(G)$
exhibit higher structural entropy, reflecting greater diversity in their
possible connectivity patterns. This measure has since found applications in
biological, chemical, and social networks.

On the other hand, time series analysis has traditionally relied on
statistical and spectral tools to characterize dynamical systems. Over the
past two decades, however, an alternative approach has emerged: representing
time series as graphs, thereby enabling the application of complex network
theory to investigate temporal patterns. Among the various methods proposed
for such transformations, the visibility graph (VG) and its variants have
become particularly popular due to their conceptual simplicity and strong
physical interpretability. In the visibility algorithm introduced by Lacasa 
\emph{et al.} \cite{Lacasa2008}, each data point $(t_i, x_{t_i})$ of
a time series is represented as a node $i$, and two nodes $i$ and $j$ are
connected if one can draw a straight line between $(t_i, x_{t_i})$ and $%
(t_j, x_{t_j})$ without intersecting any other line connecting data points 
in between.

This mapping preserves several properties of the original series
such as periodicity, fractality, and correlations within the
topological features of the resulting graph. The degree distribution,
clustering, and motif statistics of visibility graphs often reflect
fundamental aspects of the underlying dynamics, distinguishing, for
instance, chaotic, stochastic, and correlated processes. A simplified
version, known as the Horizontal Visibility Graph (HVG) \cite{Luque2009},
imposes a stricter geometric criterion for visibility, yielding greater
analytical tractability and robustness against noise. Remarkably, visibility
graphs derived from independent and identically distributed (i.i.d.) random
sequences exhibit an exponential degree distribution, whereas correlated
signals or multifractal time series deviate from this universal behavior 
\cite{Lacasa2010,Ni2009}.
 
For large time series, Lan et al.\ \cite{Lan} proposed an efficient algorithm of complexity $O(n \log n)$ to handle larger instances. Therefore, Visibility graph methods have found applications across physics, geophysics, finance, biomedicine, and climate science, providing a bridge between nonlinear dynamics and network theory \cite{Donner}. In \cite{Ren}, the concept of visibility graphs in phase space is introduced, aiming to capture more complex dynamical properties and to differentiate between chaotic and stochastic systems.

This framework enables the topological quantification of temporal complexity and the extraction of
scaling properties directly from the network structure. But what about time
series generated by spin system models? Zhao \emph{et al.} \cite{Zhao2017}
investigated networks constructed from time series of the Ising model, and Moraes 
and Ferreira \cite{Ferreira} similar results on the contact process, both yet without 
exploring the spectral properties of random matrices represented by adjacency matrices or, 
more importantly, addressing the main question of the present work: what can the number of 
spanning trees of graphs constructed from different evolutions of spin systems governed by the
traditional Monte Carlo Markov Chain (MCMC) prescribed by the Metropolis algorithm
reveal about the phase transitions of such systems?

In this paper, we propose an unconventional yet conclusive study on the
meaning of the number of spanning trees or, more precisely, its
logarithm in relation to the criticality of a classic and
paradigmatic model: the Ising model. Our goal is merely to illustrate a
method that can be readily extended to other systems. We show that the
eigenvalues of Laplacian cofactors are sensitive to the model's critical point, 
and that the eigenvalue density of adjacency matrices derived from visibility 
graphs exhibits clear deviations from those obtained for Gaussian noise. In particular, 
these spectra differ significantly from the Wigner semicircle law or any perturbation
thereof typically associated with random graphs, revealing that
the subcritical Ising model possesses a longer tail density and
supercritical Ising model shorter tails distinctive eigenvalue density,
which is also reflected in the spacing eigenvalue distribution.



\textbf{SPECTRAL SIGNATURES OF SPIN DYNAMICS THROUGH VISIBILITY GRAPHS:} Let us define
\begin{equation*}
m_{i,j} = \frac{1}{L^{2}} \sum_{k=1}^{L^{2}} s_{k}^{(i,j)}
\end{equation*}
as the magnetization per spin at the $i$-th Monte Carlo (MC) step ($i = 0, 1, 2, \ldots, N_{\mathrm{steps}} - 1$) of the $j$-th independent run ($j = 1, \ldots, N_{\mathrm{run}}$). Here, the spin variables $s_{k}^{(i,j)} = \pm 1$ correspond to those of the model chosen for testing in this work, namely the standard two-dimensional Ising model without an external magnetic field, $\mathcal{H} = -J \sum_{\langle k,k' \rangle} s_k s_{k'}$,
where $\langle k,k' \rangle$ indicates that interactions occur only between nearest neighbors. This same definition applies to other spin models evolved from random initial configurations. The evolution may use the Metropolis algorithm (as in this work) or alternative prescriptions like heat bath or Glauber dynamics \cite{Newman}, with each configuration representing a distinct run.

Several studies have shown that, during the early stages of evolution, criticality can be identified through temporal power laws \cite{Janssen1989,Huse1989,Zheng1998,Albano,Henkel}, from which all critical exponents of different models can be precisely determined \cite{Zheng1998,Albano,Silva2012,Okano}. For example, for $m_{0j}=1$, it is expected that, at $T = T_C$ and for sufficiently large systems, $m_i = \frac{1}{N_{run}} \sum_{j=1}^{N_{run}} m_{ij} \sim i^{-\frac{\beta}{\nu z}}$, where $\beta$ and $\nu$ are the standard static exponents, and $z$ is the dynamic exponent. For $T \neq T_C$, deviations from this power law are typically observed and are well described by a stretched exponential. 

On the other hand, for simulations starting from a small but fixed initial magnetization, $m_{0j} = \varepsilon \ll 1$, the system exhibits a power-law behavior characterized by a new dynamic exponent $\theta$, such that $m_i \sim i^{\theta}$ (see, for example, Refs.~\cite{Henkel2,Silva2002,Silva2013,Okano}).

This exponent can also be determined without fixing the initial magnetization--i.e., by considering random initial configurations ($T \rightarrow \infty$)--through averaging over the initial magnetizations using the total correlation $c_i = \frac{1}{N_{run}} \sum_{j=1}^{N_{run}} m_{ij} m_{0j}$, where $\langle m_{0j} \rangle = \frac{1}{N_{run}} \sum_{j=1}^{N_{run}} m_{0j} \approx 0$. In this case, the same power-law behavior is observed, namely $c_i \sim i^{\theta}$  \cite{Tome1998}.

Thus, we can construct the visibility graph for the $j$-th evolution, represented by the time series $m_{0j}, m_{1j}, \ldots, m_{N_{\mathrm{steps}}-1,j}$. The adjacency matrix of the visibility graph is defined by representing each data point $(i, m_{i,j})$ in the time series as a node $i$. Two nodes $i$ and $i'$ are said to be adjacent (or connected) if a straight line can be drawn between the points $(i, m_{i,j})$ and $(i', m_{i',j})$ without intersecting any intermediate points. The adjacency matrix $A$ is then defined as:

\begin{equation*}
a_{i,i^{\prime }}=\prod_{k=i+1}^{i^{\prime }-1}H\left( m_{i,j}+(m_{i^{\prime
},j}-m_{i,j})\cdot \frac{(k-i)}{(i^{\prime }-i)}-m_{k,j}\right) 
\end{equation*}%
where $H(x)$ is the Heaviside step function:%
\begin{equation*}
H(x)=\left\{ 
\begin{array}{ccc}
0 & \text{if} & x<0 \\ 
&  &  \\ 
1 & \text{if } & x\geq 0.%
\end{array}%
\right. 
\end{equation*}%
In other words, for node $i$ to be adjacent to node $i'$, it is necessary that $H = 1$ for all $k$ from $i+1$ up to $i'-1$. The matrix $A$ is a binary symmetric matrix with a null diagonal. It is worth noting that, although random graphs are similar to random matrices, their construction methods differ: in random graphs, with probability $p$ a node $i$ is connected to node $j$, and with probability $1-p$ they are not connected.

The eigenvalues of symmetric matrices of dimension $N$, whose entries are random variables identically distributed according to a symmetric probability density
function (PDF) with all moments finite, necessarily follow Wigner’s
semicircle law \cite{Wigner}: $\rho(\lambda) = \frac{\mathbf{1}_{\left|\lambda\right| \leq
2\sigma\sqrt{N}}}{2\pi N\sigma^{2}} \sqrt{4N\sigma^{2} - \lambda^{2}}$, where
$\sigma^{2}$ is the second moment of the matrix entries and
$\mathbf{1}_{\left|\lambda\right| \leq R}$ is the indicator function, which takes
the value 1 for $\left|\lambda\right| \leq R$ and 0 otherwise.

Although adjacency matrices $A$ of random graphs, when standardized as
$A_{s} = \frac{A - pJ}{\sqrt{Np(1-p)}}$, where $J_{ij} = 1$ for
$i,j = 1, \ldots, N$, have eigenvalues whose bulk converges (as
$N \rightarrow \infty$) to Wigner’s semicircle law,
$\rho(\lambda) = \frac{\mathbf{1}_{\left|\lambda\right| \leq 2}}{2\pi}
\sqrt{4 - \lambda^{2}}$, in the dense regime (where $p$ is constant) one
eigenvalue escapes from the bulk with a pronounced value
$\lambda(A) = Np + o(N)$ \cite{Metz,Furedi}. For the intermediate case, when
$Np$ is of the order of a few tens, the same centering and normalization yield a good fit
to the semicircle law, although small fluctuations may occur
\cite{Newman3}. Finally, in the sparse regime, $p = c/N$ with $c$
constant, the graph becomes sparse and behaves similarly to a random tree,
since its construction follows a Galton–Watson stochastic (branching)
process. In this case, the histogram of eigenvalues no longer follows the semicircle law:
there is a concentration -- or aggregation -- near zero and a flattened bulk,
accompanied by the occurrence of outliers outside this interval
\cite{Goh,Bianconi2,Albert2002}.

What can we expect from the adjacency matrices $A$ of visibility graphs
with $N_{\mathrm{steps}}$ nodes, constructed from the time series of the
magnetization evolution with $N_{\mathrm{steps}}$ Monte Carlo steps, at
different temperatures? And what about the number of spanning trees at
different temperatures? Could the behavior of this quantity reflect the
thermodynamic behavior of the system? Bianconi \cite{Bianconi2015} discussed
how Laplacian spectra relate to the thermodynamic properties of statistical
models on networks, such as Ising and percolation systems. However,
extracting such properties from graphs constructed via the visibility
representation of the magnetization time evolution -- analyzed through the lens of random matrix theory -- is both surprising and novel.

For this purpose, given an adjacency matrix $A$, one can construct another
fundamental matrix: the Laplacian matrix, defined as
\begin{equation*}
L = -A + D,
\end{equation*}
where $D$ is a diagonal matrix with elements $D_{ii} = \sum_{j=1}^{N_{\mathrm{steps}}} a_{ij}$,
and, by construction, $a_{ii} = 0$. If $A$ corresponds to the adjacency matrix
of a connected graph $G$, then an important theorem can be stated:

\textbf{Kirchhoff’s Theorem} (Matrix–Tree Theorem, see for example \cite{Harary1969}). 
All cofactors of the Laplacian matrix $L$ are equal, and each of them corresponds to the number of spanning trees of the graph $G$.

For simplicity, one typically selects the first cofactor, so that the number of spanning trees of $G$ is given by
\begin{equation*}
\tau(G) = \det(L') = \prod_{i=1}^{N_{\mathrm{steps}} - 1} \lambda_i(L'),
\end{equation*}
where $\lambda_1, \lambda_2, \ldots, \lambda_{N_{\mathrm{steps}} - 1}$ are the
eigenvalues of the symmetric matrix
\begin{equation*}
\left[L'\right]_{ij} = 
\begin{cases}
\displaystyle \sum_{k=1}^{N_{\mathrm{steps}}} a_{i+1,k}, & \text{if } i = j, \\[1em]
-\,a_{i+1,j+1}, & \text{otherwise.}
\end{cases}
\end{equation*}

The distinctive aspect of our work is that we employ the spectra of random matrices
derived from adjacency matrices not merely to study graph properties, 
but to reveal critical features of systems whose time evolutions are sampled over very few Monte Carlo (MC) steps 
(typically around $N_{\mathrm{steps}} = 100$). 
From a numerical standpoint, we propose to use a metric---the \textit{structural entropy}---normalized by the number of MC steps minus one:
\begin{equation}
\begin{array}{lll}
N_{t} & = & \frac{1}{(N_{\mathrm{steps}} - 1)} \ln \tau(G) \\[0.7em]
& = & \frac{1}{(N_{\mathrm{steps}} - 1)} \sum_{i=1}^{N_{\mathrm{steps}} - 1} \ln \lambda_i(L').
\end{array}
\label{Eq:Number_of_trees}
\end{equation}



\begin{figure*}[t!]
    \centering
    \includegraphics[width=0.9\textwidth]{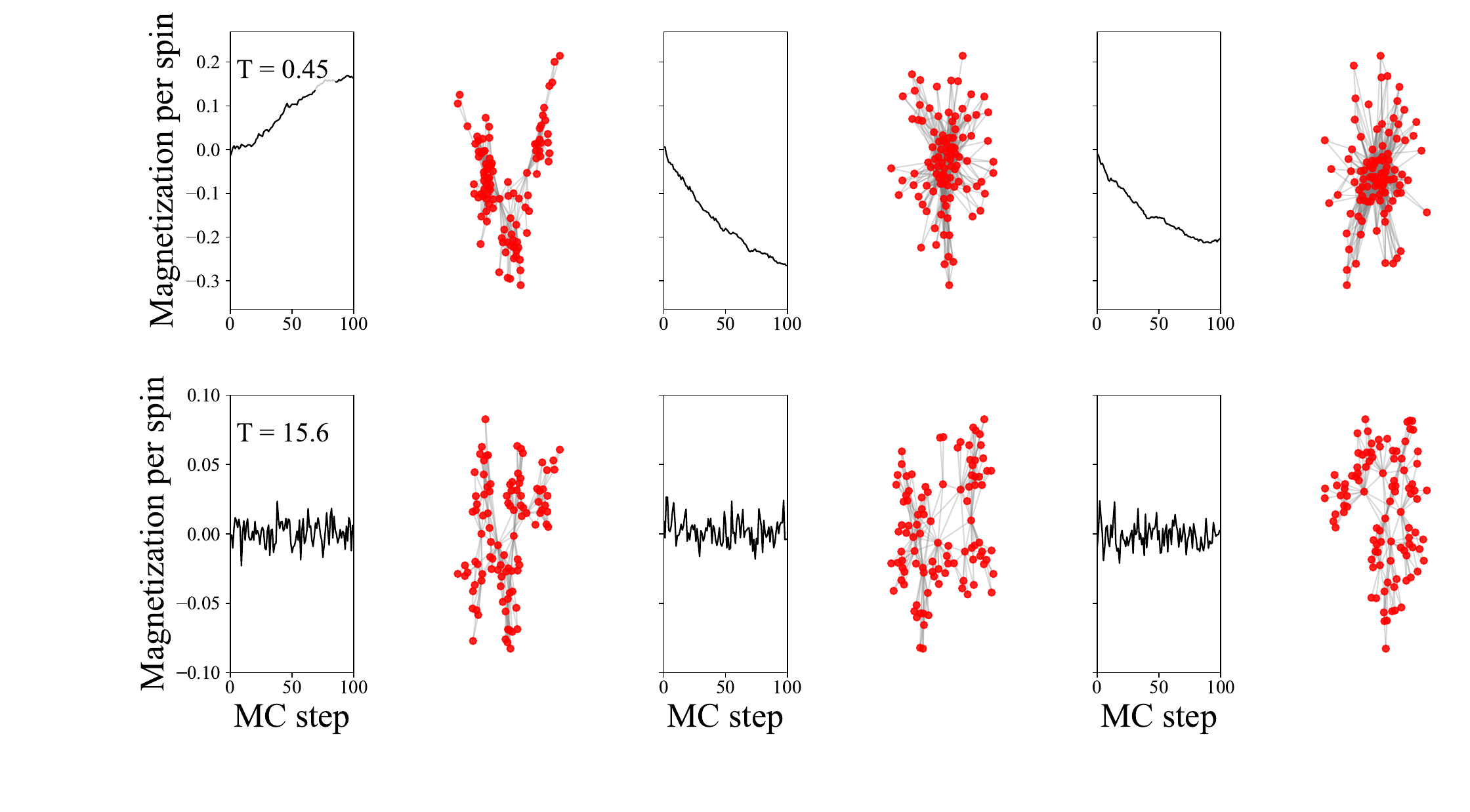}
    \caption{Time evolution of magnetization for three different random seeds
    at two temperatures, $T=0.45$ and $T=15.6$. The corresponding visibility
    graphs, constructed from these series, reflect distinct spectral
    properties in the random adjacency matrices derived from them.}
    \label{Fig:examples}
\end{figure*}

\textbf{RESULTS:}
We begin our results section by presenting examples of visibility graphs constructed from time series of the magnetization per spin of the Ising model, evolved via MCMC simulations using the Metropolis algorithm. The time evolutions of the system for two different temperatures $T = 0.45$ and $T = 15.6$, each for three distinct random seeds, are shown in Fig.~\ref{Fig:examples}.

At low temperature ($T = 0.45$), below the critical one $T_{C} = \tfrac{1}{2} \ln(1 + \sqrt{2})$ the strong spin-spin correlations in the ordered Ising system give rise to slow, collective fluctuations of the magnetization. The corresponding visibility graphs display hubs, higher clustering, and a more heterogeneous degree distribution, reflecting the strong temporal correlations. In contrast, at high temperature ($T = 15.6$), the magnetization fluctuates around zero with almost no temporal correlation. The visibility graphs are more homogeneous, with a narrow degree distribution and low clustering, resembling a random network, which captures the increased stochasticity of the dynamics.

\begin{figure}[h]
\centering\includegraphics[width=0.5%
\textwidth]{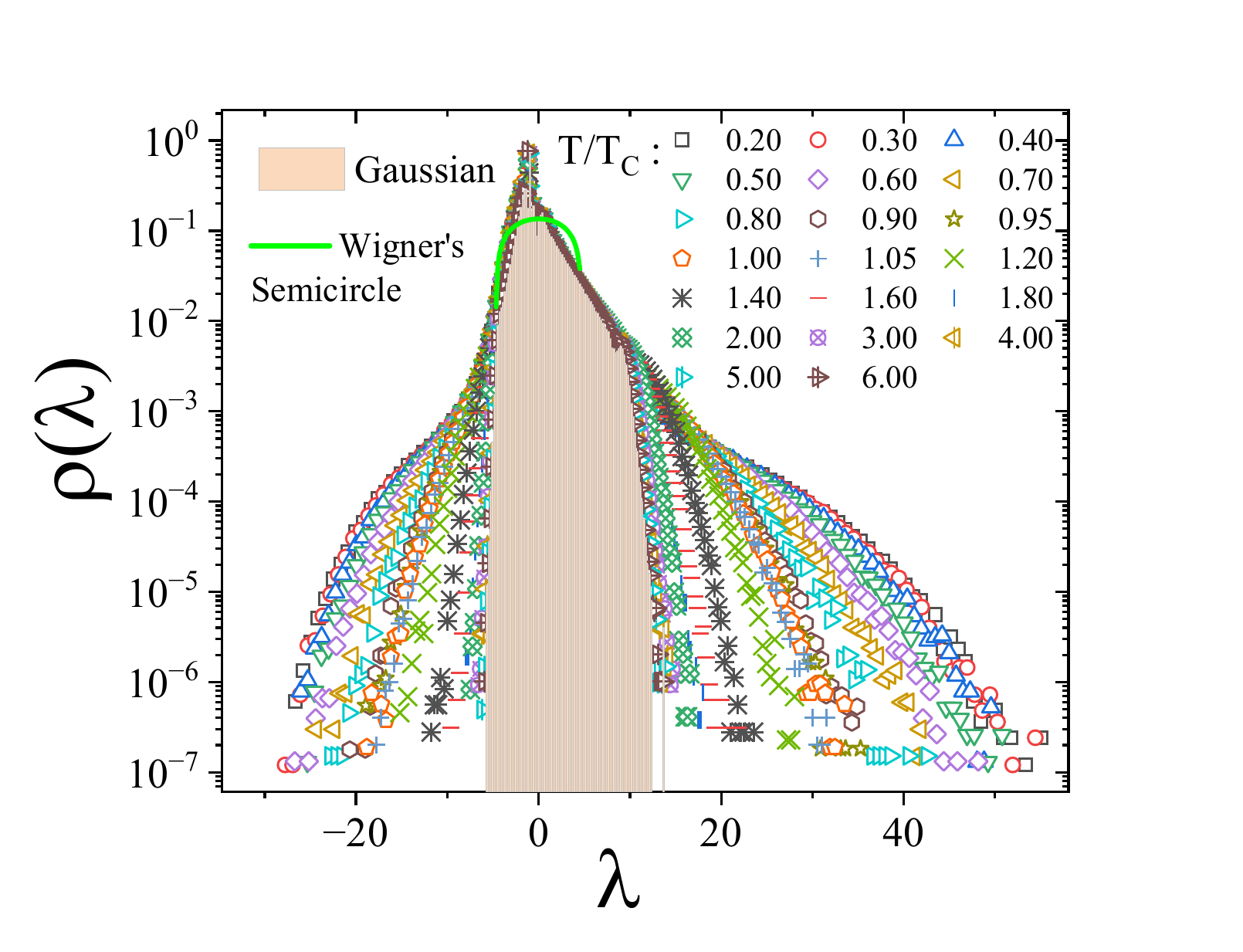}
\caption{Density of eigenvalues of adjacency matrices constructed from visibility 
graphs of magnetization time series at different temperatures. At high temperatures, 
the densities resemble those of Gaussian noise visibility graphs, shown with bars for 
comparison. At lower temperatures, the distributions exhibit heavier tails.}
\label{Fig:density_of_eigenvalues}
\end{figure}

We used $L=128$ and $N_{steps}=100$ MC steps, which are the parameters used throughout this paper. We then focus on the density of eigenvalues of the adjacency matrix $A$. For this purpose, we prepared an ensemble of $N_{run}=10^{5}$ matrices constructed from different time evolutions, yielding a total of $10^{7}$ eigenvalues to build the histograms. For each studied temperature from $T=0.2T_{C}$ until $T=6T_{C}$, we constructed the corresponding histogram of eigenvalue densities (see Fig.~\ref{Fig:density_of_eigenvalues}).

\begin{figure}[h]
\centering\includegraphics[width=0.5\textwidth]{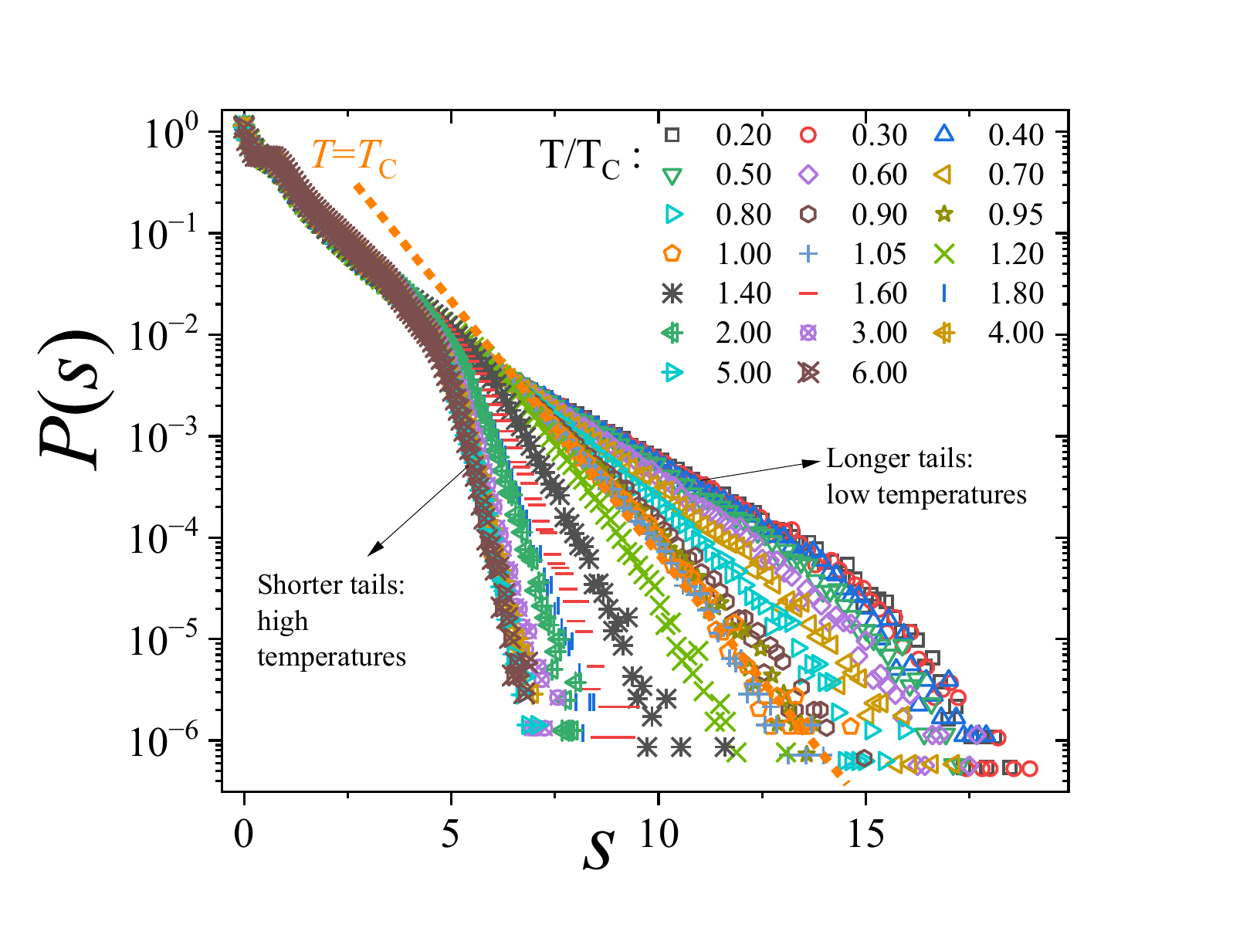}
\caption{Eigenvalue spacing distribution. The same temperatures used for the density plots are shown here. Shorter tails are observed at higher temperatures, while longer tails appear at lower temperatures, consistent with the behavior seen in the eigenvalue density distributions.}
\label{Fig:Spacing}
\end{figure}

The density of eigenvalues was obtained from adjacency matrices constructed via visibility graphs of magnetization time series at different temperatures. At high temperatures, the distributions resemble those of Gaussian noise visibility graphs, shown with bars for comparison. At lower temperatures, the distributions exhibit heavier tails. The green continuous curve represents the Wigner semicircle law. This behavior is also reflected in the eigenvalue spacing, defined as $s_i = \frac{\lambda_i - \lambda_{i-1}}{D}$, where $D = \frac{1}{(N_{\mathrm{steps}}-1)} \sum_{i=2}^{N_{\mathrm{steps}}} (\lambda_i - \lambda_{i-1}) = \frac{\lambda_{N_{\mathrm{steps}}} - \lambda_1}{N_{\mathrm{steps}}-1}$ is the average spacing. Using the same ensemble of matrices, we constructed histograms of eigenvalue spacings at the same temperatures, which exhibit patterns similar to those observed for the eigenvalue densities, as shown in Fig.~\ref{Fig:Spacing}.

Shorter tails are observed at higher temperatures and longer ones at lower temperatures, consistent with the behavior of the eigenvalue density. The case $T = T_{C}$ is highlighted for reference. Qualitatively, the eigenvalue density of visibility graphs constructed from the temporal evolution of magnetization appears to reflect the system’s thermodynamics. Similar spectral signatures have been reported in the analysis of correlation matrices (see, e.g., \cite{RMT2023,RMT2023-2,RMT2023-3,Seligman3,Vinayak2014,Biswas2017}). We next examine whether the logarithm of the number of spanning trees, defining the structural entropy, can serve as a thermodynamic indicator. In particular, we test whether the number of spanning trees responds to the model’s phase transition.

\begin{figure}[h]
\centering
\includegraphics[width=0.5\textwidth]{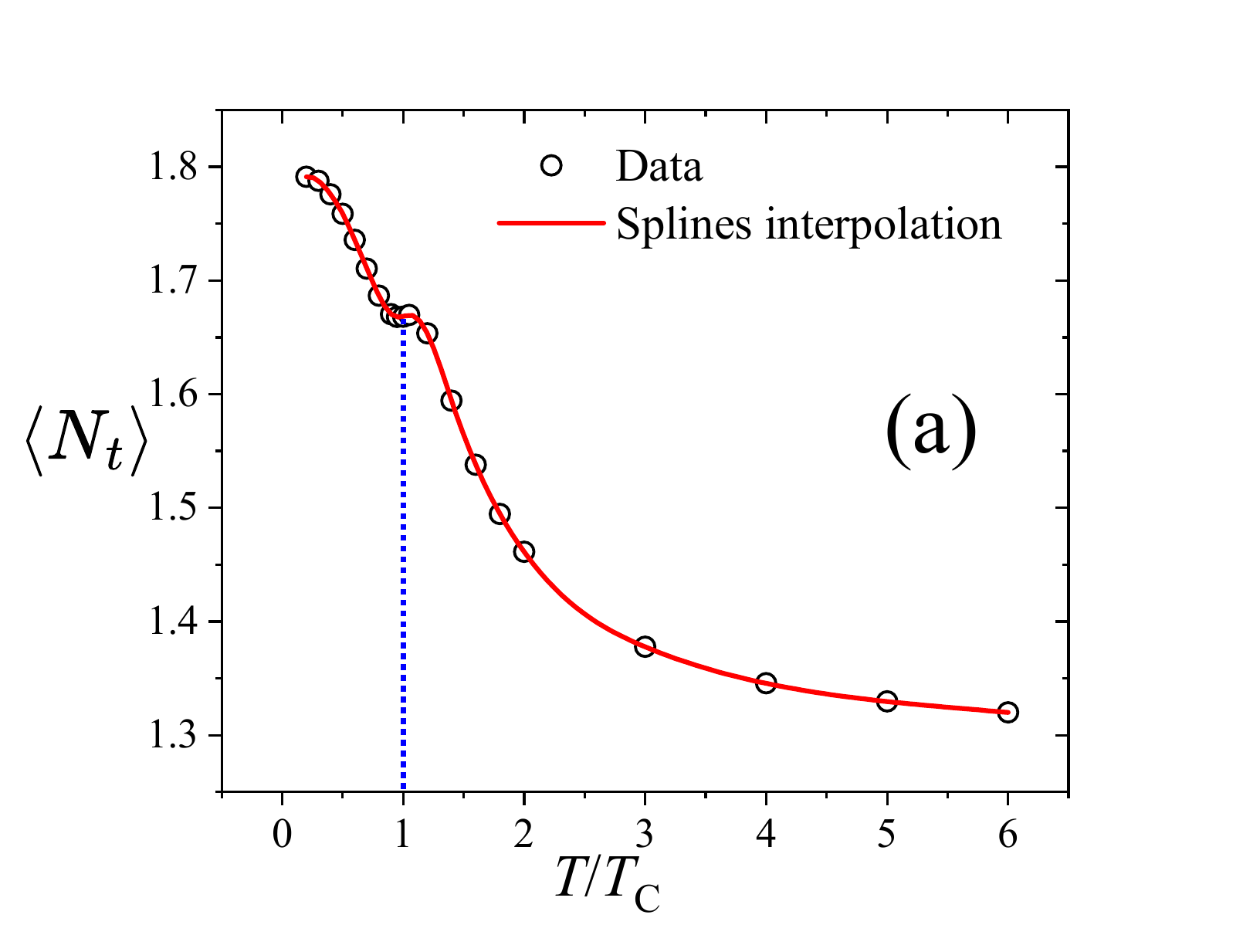}\\
\includegraphics[width=0.5\textwidth]{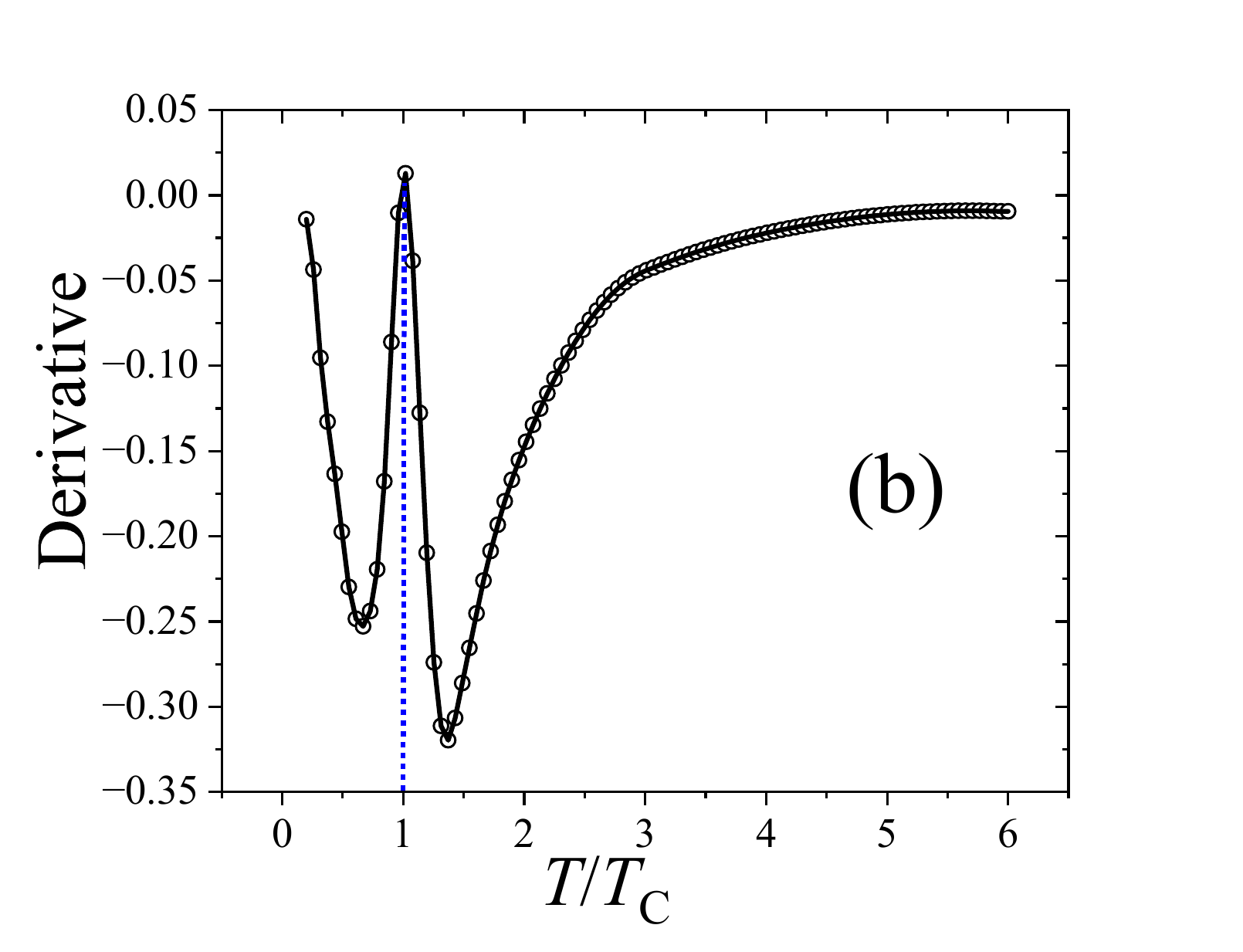}
\caption{Fig.~(a) Average of the logarithm of the number of spanning trees (structural entropy) per MC step for different temperatures. Fig.~(b) Derivative of the quantity shown in Fig.~(a) for the same temperatures. Both quantities appear to signal a phase transition at $T \approx T_{C} = \frac{1}{2} \ln(1 + \sqrt{2})$, indicating that the critical point corresponds to the maximum of the structural entropy.}
\label{Fig:Phase_transition}
\end{figure}

Using Kirchhoff’s theorem and recalling that the number of spanning trees of a connected graph is a fundamental combinatorial measure of connecting all nodes without redundancy, we compute the number of spanning trees for the different temperatures considered in the analysis of eigenvalue densities and spacing distributions. We then evaluate $\langle N_{t} \rangle$, the structural entropy, defined as the ensemble average of the quantity in Eq.~\ref{Eq:Number_of_trees}. The resulting dependence of $\langle N_{t} \rangle$ on $T/T_{C}$ is shown in Fig.~\ref{Fig:Phase_transition}(a).

It is interesting to observe that this quantity exhibits a small plateau near $T/T_{C}=1$. In this situation, it is convenient to calculate the derivative of $\left\langle N_{t}\right\rangle$ with respect to $\tau = T/T_{C}$. The derivative is plotted as a function of temperature in Fig.~\ref{Fig:Phase_transition}(b). To obtain a smooth derivative, we first interpolate $\left\langle N_{t}\right\rangle$ as a function of $\tau$ using splines and then differentiate the resulting curve. A pronounced peak is observed at $T/T_{C} \approx 1$, corroborating the behavior qualitatively described for the density of eigenvalues and spacing distribution of $A$. This indicates that the critical point is associated with the maximum of the structural entropy.

It is important to mention that these results are similar to those obtained from the density of eigenvalues of correlation matrices, which also reflect comparable phenomena. In that case, analyzing the fluctuations (first and second moments) of the eigenvalues as a function of $T/T_{C}$ allows one not only to capture these effects but also to identify chaotic transitions and systems in the mean-field regime \cite{Silva2025}. 

Last but not least, we would like to highlight the power of the maximum entropy principle \cite{Jaynes}, which we employ here to show that the structural entropy, constructed from the number of spanning trees obtained from the time series of the magnetization of the Ising model evolved via the Metropolis MCMC algorithm, reaches its maximum at the critical temperature. The versatility of this approach can be appreciated across various contexts, ranging from estimates of the Higgs and axion masses in elementary particle theory \cite{Higgs,Axion} to quantitative analyses in tropical forest ecology \cite{Pos}. In this work, we introduce a novel application that may in the future be extended to other time series from diverse complex systems. 



\textbf{CONCLUSIONS:} To summarize, we propose a novel analysis of visibility graphs constructed from the time evolution of magnetization shows that the number of spanning trees, or simply its logarithm, provides a nonconventional indicator of the thermodynamics of spin systems and captures the phase transition of the Ising model, similarly to the behavior observed in the histograms of eigenvalues and eigenvalue spacings of the adjacency matrices ---a perspective that, to our knowledge, has not been explored before. It is, of course, necessary to reinterpret the meaning of this “spectral thermodynamics,” particularly regarding the connection between criticality and the maximum of the structural entropy (also referred to as tree entropy when divided by the number of nodes in the thermodynamic limit; see, for example, \cite{Lyons}). 

Nevertheless, this concept appears to be a key element in linking the system’s critical behavior with the structure of the graph generated from the time series. This study suggests that similar analyses could be extended to other models in Statistical Physics, and even to other physical phenomena represented by time series of relevant observables. It is important to mention that Schnakenberg \cite{Schnakenberg} applies Kirchhoff’s ideas to a master equation, interpreting the transition rates as analogous to resistors in an electrical network. Our time series is derived from the evolution of order parameters governed by dynamics defined through the master equation under a specific condition — the transition rates satisfy detailed balance. This framework opens promising avenues for future investigations. 


\textbf{ACKNOWLEDGEMENTS:} RDS, HAF, SG, and AA thank CNPq (Conselho Nacional de Desenvolvimento Científico e Tecnológico) of Brazil for financial support under grants 304575/2022-4, 405508/2021-2,  309560/2025-0, and 307317/2021-8, respectively. RDS also thanks Prof. M. J. de Oliveira (IF-USP) for his valuable insights.

\end{document}